\documentstyle[12pt,epsfig]{article}
\newcommand{\ba}{\begin{eqnarray}}
\newcommand{\ea}{\end{eqnarray}}
\newcommand{\bb}{\begin{thebibliography}}
\newcommand{\eb}{\end{thebibliography}}
\newcommand{\bi}[1]{\bibitem{#1}}

\newcommand{\ci}[1]{\cite{#1}}
\newcommand{\lab}[1]{\label{#1}}

\begin{document}

 \begin{center}
{\large \bf
  "Form of analyzing power and the determination of
   the basic parameters of hadron scattering amplitude"}

\vspace{10mm}

 S.B.Nurushev$^{1}$,
 O.V.Selyugin$^{2}$,
 M.N.Strikhanov$^{3}$        \\
{\it $^{1}$Institute for High Energy Physics, 142284 Protvino, Moscow   \\
$^{2}$BLTPh, JINR, Dubna, Russia       \\
$^{3}$Moscow Engineering Physics Institute, Kashirrskoe Ave.31, 115409
   Moscow, Russia  }\\
\end{center}


\begin{abstract}
\begin{center}
\begin{minipage}{11cm}
   The determination of  magnitudes of  basic parameters
   of the high energy
   elastic scattering amplitude are examined
   at small momentum transfers with taking account
  of the Coulomb-hadron interference effects.
\end{minipage}
\end{center}

\end{abstract}

      An actual problem of the modern  elementary
particles physics,  the research of strong interaction processes at large
distances and  high energies, is considered in the framework of different
approaches by using various models of the structure of hadrons and the
dynamics of their interactions.
	The diffraction scattering cannot yet be described quantitatively
in the framework  of the perturbative QCD. Therefore, it is necessary
to apply different models  which can describe the hadron-hadron
interaction at large distances.
The research  of elastic  scattering requires the knowledge
of properties of the pomeron,
the object determining  the interaction of hadrons in this  range.
In this case  the study of the structure  and  spin  properties of
both the hadron  and the pomeron  acquires  a  special  role.

Now we recognize that the research of
the pomeron exchange
requires not only a pure elastic process but also many
physical processes involving electroweak boson exchanges.
There are two approaches to the pomeron, the "soft" pomeron built of
multiperipheral hadron exchanges and a more current perturbative-QCD
"hard" pomeron built of the gluon-ladder.
     The "soft" pomeron dominates in high energy hadron-hadron
diffractive reactions while the "hard" Pomeron dominates in high energy
$\Upsilon-\Upsilon$ scattering \ci{bj}
and determines the small x-behaviour of deep inelastic structure functions
and spin-averaged gluon distributions.

         The "corner stone" for many models of the Pomeron is the power
of the total cross sections growth.
The "soft" pomeron of the standard form with
$ \alpha_{pom}(0)=1 + \epsilon$
was introduced in \ci{lan1}.
 The observed growth of inelastic cross sections and the
multiplicity match this idea.
The perturbative QCD leading log calculation of the gluon ladder diagrams
gives the following result \ci{kur}:
$\   \epsilon = 12 \  \alpha_{s}/ \pi \  \ln{2} \sim 0.5  $.
This "hard" pomeron is not yet observed experimentally. Really,
the new global QCD analysis of data for various hard scattering processes
leads to the small x behaviour of the gluon structure function
 determined by the "hard" pomeron contribution   \ci{lai}
$            g(x) \sim 1/x^{1+\epsilon} $
with $\epsilon = 0.3$.

    There exist many discussions about the energy dependence
of the elastic and total cross section in  hadron-hadron
scattering \ci{blois,mart}.
Essential
uncertainty in the determination of the values of $\sigma_{tot}$
has been shown in \ci{sel1}.
Now we have a large discussion about the value of $\sigma_{tot}$
at $\sqrt{s} = 1.8 \ TeV$.
In \ci{72} it has been found that, at $\sqrt{s}=1800 \ GeV$,
 $\sigma_{tot}=72.2 \ mb$.
Recent results of the CDF Collaboration \ci{CDF}
are
$ (1+\rho^2) \sigma_{tot}=62.64 \pm 0.95 \ (mb)$ at $\sqrt{s}=546 \ GeV$,
$ (1+\rho^2) \sigma_{tot}=81.83 \pm 2.29 \ (mb)$ at $\sqrt{s}=1.8 \ TeV$.

    Let us examine the future  $pp2pp$ experiment at $\sqrt{s} =500 \ GeV$,
    as an example.
  The differential cross
  section and spin parameters $A_N$  are defined as
\begin{eqnarray}
  \frac{d\sigma}{dt}= \frac{2 \pi}{s^2}(|\phi_1|^2+|\phi_2|^2+|\phi_3|^2
   +|\phi_4|^2+4|\phi_5|^2), \lab{dsth}
\end{eqnarray}
\begin{eqnarray}
  A_N\frac{d\sigma}{dt}&=& -\frac{4\pi}{s^2}
                 Im[(\phi_1+\phi_2+\phi_3-\phi_4) \phi_5^{*})],  \lab{anth}
\end{eqnarray}
 in the framework of the usual helicity representation.

 Elastic differential cross section will be regarded as having $2 \%$
  statistical errors.
  Now, in the fitting procedure we take into account
  the standard supposition for high energy elastic hadron scattering
  at small angles: the simple exponential behavior
  with slope $B$ of imaginary and real parts of the scattering amplitude;
  the hadron spin-flip amplitude does not exceed $10 \% $
  of the hadron spin-non-flip
  amplitude.

   The differential cross section was calculated by using \ref{dsth}
  (  $ \sigma_{tot} =63.5 \ mb ;  \ \ \ \rho=0.15;
    B=15.5 \ GeV^{-2}; $  at
 $  150 $ points,  from $t_0 = 0.00075$ with $\Delta t = 0.00025 \ GeV^{2}$),
 and
   then it was put through a special random process by using $2 \% $ errors.

   After that the obtained "experimental" data are fitted (see Fig. 1).
  The systematical errors are taken into account as free parameters $n$,
  by which the fitting curve is multiplied.
    The most important value, which influences the error of the magnitude of
  $\sigma_{tot}$, is the normalization coefficient
  of the differential cros section. Its small errors lead to significant
  errors in $\sigma_{tot}$.


\begin{figure}
\centering
\mbox{\epsfysize=70mm\epsffile{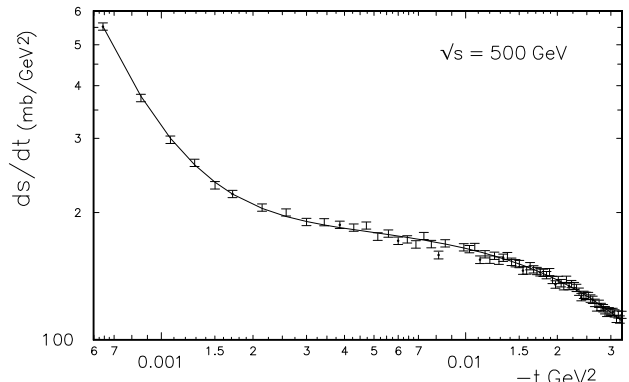}}

\vspace{-10mm}

\caption{The calculated $d \sigma / dt$ at $\sqrt{s} = 500 \ GeV$ }
\label{F1}
\end{figure}


So, for the fixed $n=1$ we obtain
  $\sigma_{tot}=63.54 \pm 0.12 \ mb$ and with free $n=1.04 \pm 0.04$
  $\sigma_{tot}=63.6 \pm 1.3 \ mb$,
    and we see that the normalization of experimental data is the most
  important problem for the definition of $\sigma_{tot}$.
  That factor  reduces polarization effects as it is
  represented as the ratio of polarization
  and unpolarization experimental data.
  So, maybe, these data help us to the determine the total cross section.
  In any case, these experiments are very important for the determination of
   the hadron spin-flip amplitude also.
  Now the diffraction processes play
   significant role in the researches of modern accelerators. There are
   some models which predict  nondecreasing spin effects at superhigh
   energies in the diffraction reactions.
  The unknown  magnitude
   of hadron spin-flip is a serious bound to use the Coulomb-nuclear effect
   for the determination of the beam polarization at future colliders.

  One important task of the $pp2pp$ experiment at RHIC is to measure
 the analyzing power, $A_N(t)$, for elastic $pp$-scattering
  in  the CNI region.
 To study this, it was looked at the performance of the apparatus
 to reconstruct an input $A_N(t)$ . The collision energy was taken as
 $\sqrt{s} = 500 \ GeV$, the beam polarization was set to 70$\%$, the running
 luminosity is assumed to be $2\times 10^{29} cm^{-2}\times s^{-1}$.
 For simulation of the
 "left-right" analyzing power, the simple form of $A_N(t)$ an given by
 equation (\ref{anth}) was applied with the usual high energy supposition
  at small momenta transfer
  for hadron spin-flip amplitudes
  with the slope of the hadron
  spin-flip amplitude equal to the slope
  of the hadron spin-non-flip amplitude
  without the kinematic parameter $\sqrt{|t|}$.

   We use the conventional helicity amplitudes $\phi_i , i=1,\ldots 5$
  as defined in \cite{lead} and assume the addition of the hadronic and
  electromagnetic amplitudes
   ($\Phi_i = \phi^{h}_i +e^{i\delta}\phi^{e}_i , i=1\ldots 5$)
  with            the Coulomb phase-shift $\delta_i$  and make the usual
  approximations $\phi^{h}_1 = \phi^{h}_3$ and $\Phi_2$ , $\Phi_4$
  being negligible at high
  energies and small t.
   Our model calculations for $A_N$ are shown in Fig.2.

     Then we check the presence of the hadron spin flip amplitude
  on our fitting
   procedure of $ds/dt$. We find that the hadron spin-flip amplitude
   that does not exceed $10 \%$ of the hadron non-spin-flip amplitude,
   without
   kinematical factor of $t$, does not change the result of fit of $ds/dt$.


\begin{figure}
\centering
\mbox{\epsfysize=70mm\epsffile{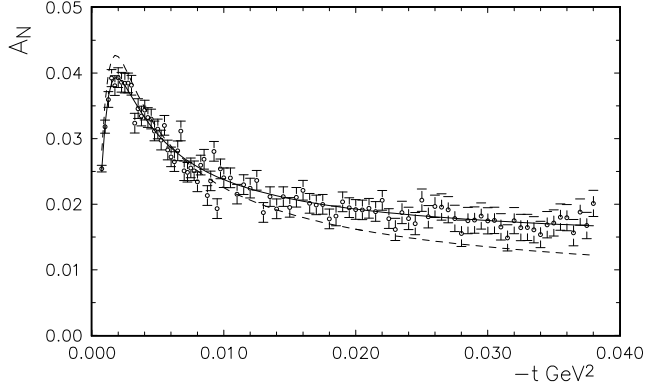}}

\vspace{-10mm}

\caption{The calculated $A_{N}$ }
\label{F1}
\end{figure}

\vspace{5mm}

   As usual, we assume that the hadron spin-flip
  amplitude is a slowly varying function of t apart the kinetic factor
  and, following \cite{bl-kor}, we parametrize it as
\begin{eqnarray}
   \phi^{h}_{5}= \frac{\sqrt{|t|}}{m} (\rho k_2 + i k_1) Im \phi^{h}_{1}. \lab{sf2}
\end{eqnarray}
  where $\rho$,$k_1$,$k_2$ are slowly changing functions of s.
  The coefficients $k_1$ and $k_2$ are the ratio of the real and
  imaginary parts of the spin-flip to spin-non-flip amplitudes
  without the kinematic factor $\sqrt{|t|}$.
 They are related to
 $R$ and $I$ in paper \cite{akth} as
\begin{eqnarray}
  I = Im \phi^{h}_{5}/(\sqrt{|t|} Im \phi^{h}_{1}) = k_1; \ \ \
	 R = Re \phi^{h}_{5}/(\sqrt{|t|} Im \phi^{h}_{1}) = \rho k_2
\end{eqnarray}

 As a result, $A_N$ can be written as:
\begin{eqnarray}
 -  \frac{A_N}{8 \pi P_{B} } \frac{d\sigma}{dt}
   =-Im \phi^{h}_{1} \frac{\alpha}{m \sqrt{|t|}} (\frac{\mu-1}{2}-k1)
          + \frac{\sqrt{|t|}}{m} \rho [Im \phi^{h}_{1}]^{2}
	 (k_2-k_1)                                          \lab{kil}
\end{eqnarray}

 For example, when the phase between the hadron spin-non-flip
  and spin-flip amplitudes are equal at small transfer momenta,then
  $k_1=k_2$, and there is no term in the polarization depending only on the
  hadron amplitudes. In that case we clearly see that
  we have an additional contribution to the analyzing power coming from
  the imaginary part of the hadron spin-flip amplitude
  and, which is most important, having the same form as the basic term
  of the Coulomb-nucleon interference.
    This contribution has the same
  effect as the error of the beam polarization, i.e. if the imaginary
  part of the hadron spin-flip amplitude (without the kinematic parameter)
  is only $5\%$ of the imaginary part of the hadron non-flip amplitude,
  we have the $5\%$ error in the definition of the beam
  polarization from the Coulomb-nucleon interference effect.

\begin{table}[htb]
 \begin{center}
 \caption{Fit of $A_N$ }
 \label{t3}
 \vspace{1mm}
 \begin{tabular}{|c|c|c|c|c|c|} \hline
 $N$  & $\sigma_{T} \ mb $ & $\rho$  & $k_1$ & $ k_2$ &$ n_2$ \\
 $ A_1$ & $63.5 \pm 3.4$ &  $0.14 \pm 0.08$ & $-0.005 \pm 0.07$
                        & $0. \pm 0.05$   & $ 1. $ -fix  \\
 $ A_2$ & $63.46 \pm 3.8$ &  $0.14 \pm 0.15$ & $ 0.1 \pm 0.06$
                             & $0.1 \pm 0.1  $  & $ 1.13 \pm 2.4 $  \\
 $ B_1$ & $63.5 \pm 3.8$ &  $0.14 \pm 0.09$ & $0.095 \pm 0.07$
                        & $0.14 \pm 0.11  $ & $ 1. $ -fix  \\
 $ B_2$ & $62.7 \pm 4.$ &  $0.13 \pm 0.3$ & $ 0.05 \pm 6.3$
                             &$ 0.1 \pm 5.6  $      & $ 0.93 \pm 3.7 $  \\
 $ C_1 $& $63.4 \pm 3.6$ &  $0.14 \pm 0.09$ & $0.095 \pm 0.07$
                        &$ -0.14 \pm 0.11 $  & $ 1. $ -fix  \\
 $ C_2$ & $63.5 $ -fix &  $0.15 $-fix & $ 0.1 \pm 0.015$
                             &$ -0.14 \pm 0.011 $ & $ 1. $ - fix  \\
 $ C_3$ & $63.9 \pm 1.83 $ &  $0.01 \pm 0.05 $ & $ 0.06 \pm 0.037$
                             &$ -0.05 \pm 0.035 $ & $ 1. $ - fix  \\
  \hline
 \end{tabular}
 \end{center}
\end{table}

\begin{figure}
\begin{flushleft}
\mbox{\epsfysize=50mm\epsffile{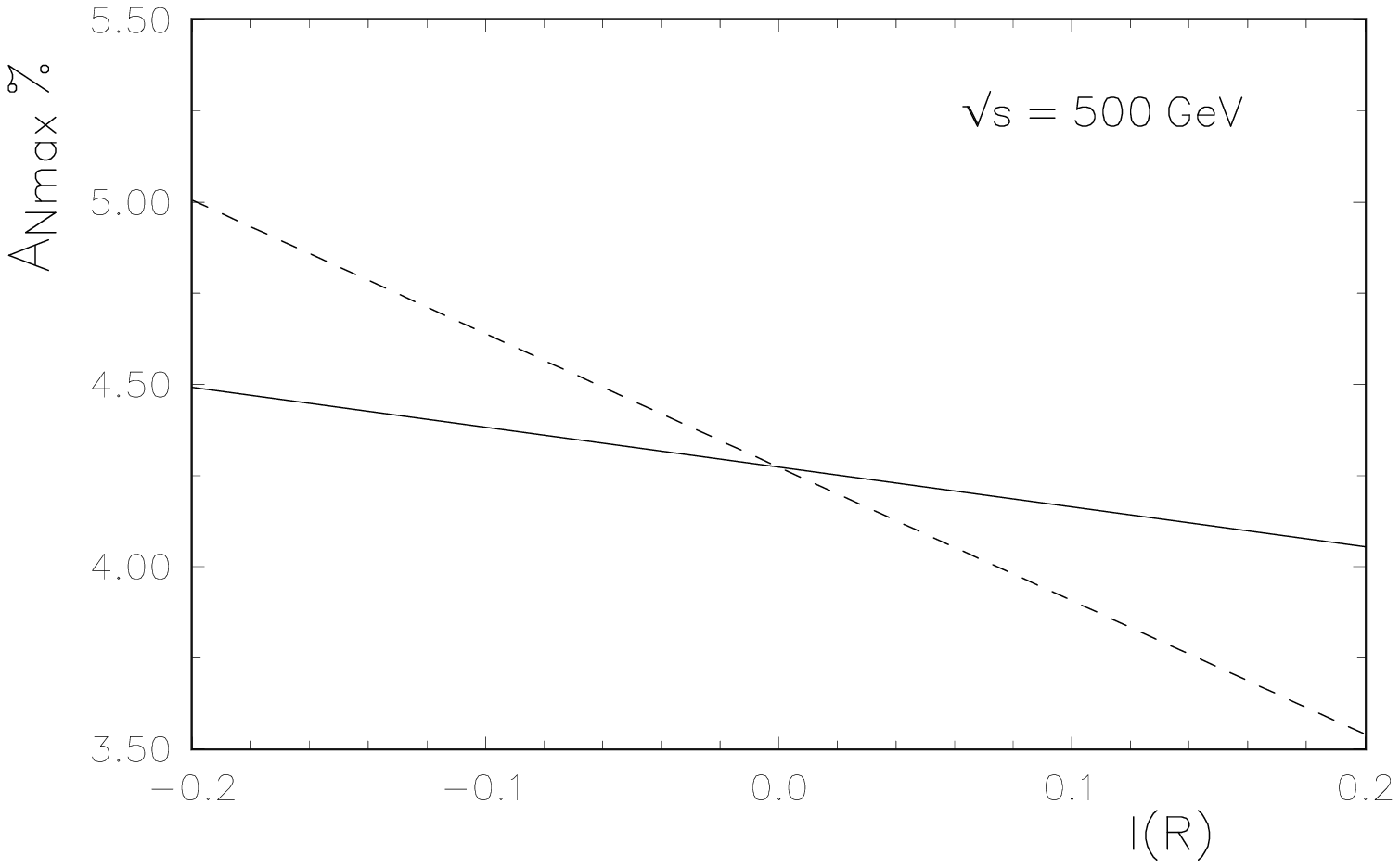}}
\end{flushleft}

\vspace{-5.83cm}

\begin{flushright}
\mbox{\epsfysize=50mm\epsffile{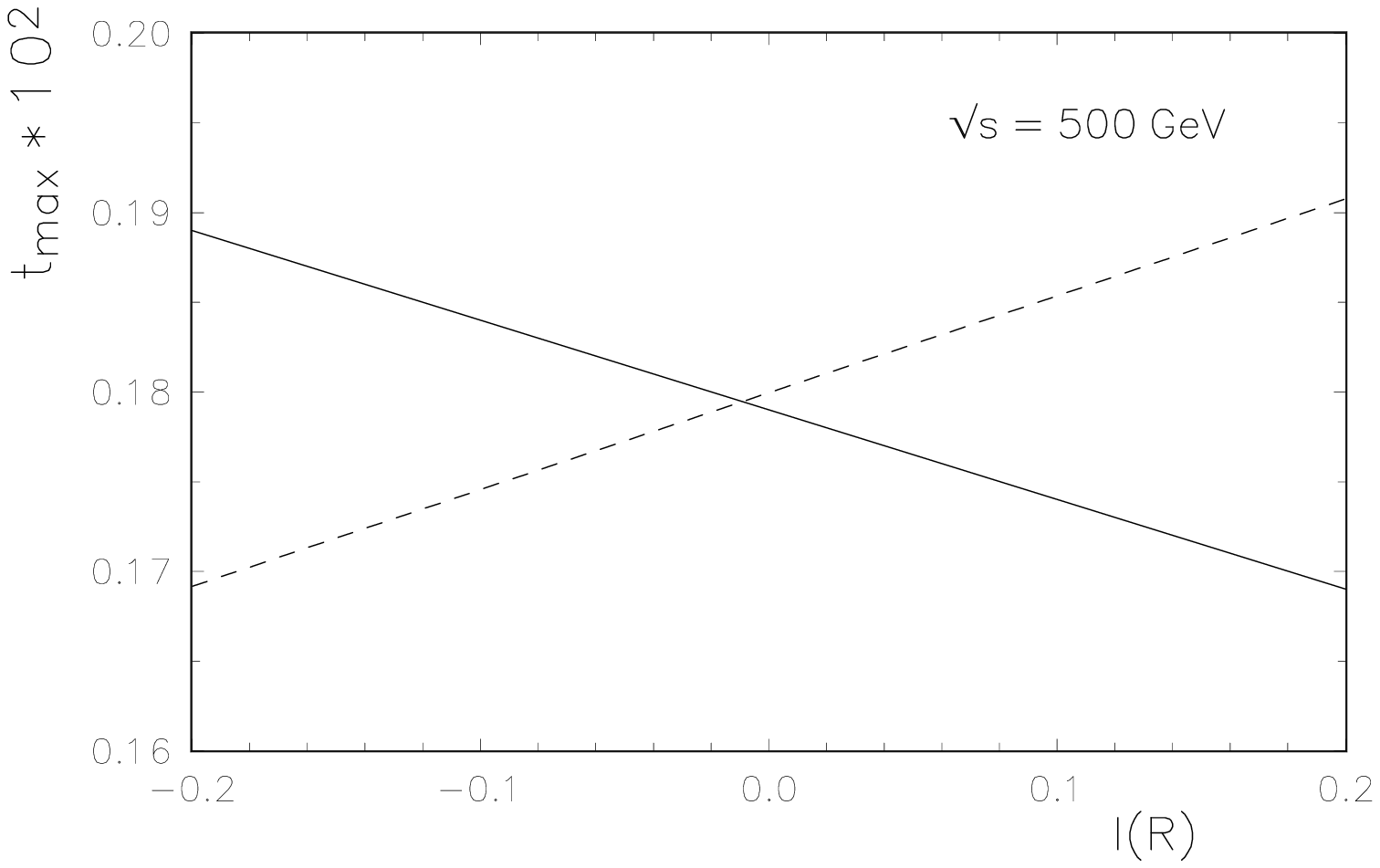}}
\end{flushright}

\vspace{-.5cm}

\begin{minipage}{5.7cm}
{Fig. 3 $A_N$ at the point of maximum of CNI at $\sqrt{s}=500 \ GeV$.
  (solid and dashed lines -
   the contributions from $Re F_h^{+-}$ and $Im F_h^{+-}$) }
\end{minipage}

\vspace{-25mm}
\begin{flushright}
\begin{minipage}{5.7cm}
Fig. 4 The position of the maximum of CNI
  at $\sqrt{s}=500 \ GeV$.
  (solid and dashed lines - the contributions from $Re F_h^{+-}$ and $Im F_h^{+-}$)
\end{minipage}
\end{flushright}
\end{figure}

\vspace{.5cm}

  This fit in case $A_N$ is presented in Fig.2 by a solid line.
 The dashed line in  Fig.2 shows the CNI effect without the
 hadron spin-flip amplitude.

  Let us made the fit for both the data on the differential cross section and
  analyzing power. The new fit gives a slight decrease in the error of
  $\sigma_{tot}$ approximately by $10\%$.
  But the determination of
  the magnitude of real and imaginary parts of the hadron spin-flip
  amplitude becomes  three time more accurate.

      As the contribution of that hadron spin-flip amplitude to the
   maximum of CNI effect is $8 \%$, it gives  the uncertainty in
   the definition of maximum $A_N$ in the region of CNI effect with $2 \%$
   errors.

   The dependence of the additional contribution to CNI effect
    on the imaginary and real parts of the hadron-spin flip
    amplitude at $\sqrt{s} = 500 \ GeV $ is shown in Fig. 3.
   It is clear that this contribution has the same sign for both parts.
   The reason is in fact that the basic contribution $k_1$
    gives in first term of $A_N$ which have practically the same
    form as usual CNI effect. But at other points of $t$, this contribution
    is  different in sign for imaginary and real parts of the hadron
    spin-flip amplitude. This is reflected in the second term
    of (\ref{anth}) that contains $(k_2-k_1)$.

 Now we turn to the discussion of difficulties which can be encountered in
 applications of a new scheme to  experiments. Mostly they are systematic
 errors in the beam polarization.
  If we take into account the systematic errors of the beam polarization
   as $n_2$ in variants $B$ and $C$, by which we multiply our fitting curve,
   we find that the coefficients $k_1$ and $k_2$ cannot be found
   separately. In this case we can find only  some ratio
   of the real to imaginary parts of the hadron spin-flip amplitude
   \cite{bl-kor}. But the point of maximum $A_N$ is  independent
   of the systematic errors.

 This value is tightly connected
   with the magnitude of $\sigma_{T}$ and other parameters.
   The dependence of $t_{max}$ on $k_1$ and $k_2$ is shown in
   Fig. 4. It can be seen that this dependence is  different in sign
   for real and imaginary parts of the hadron spin-flip amplitude.
    Our $10 \%$ hadron spin-flip amplitude leads to a small exchange
    of the point of maximum $A_N$. We obtain for the clear ($k1=k2=0$) CNI
    effect ar $\sqrt{s} = 500 \ GeV$
    that $-t_{max} = 1.179 10^{-3} \ GeV^{2}$. For the case when
    $k_1 = 0.1$ and $k_2= -0.15$, we obtain
      $-t_{max} = 1.86 10^{-3} \ GeV^{2}$ \\

{\bf Acknowledgements}

We would like to thank  V.Emelyanov,
and B. Nicolescu   for useful discussions.

\end{document}